\documentclass[manuscript]{acmart}
\usepackage{amsfonts}
\usepackage{graphicx} % Required for inserting images
\usepackage{enumitem}

\AtBeginDocument{%
  \providecommand\BibTeX{{%
    \normalfont B\kern-0.5em{\scshape i\kern-0.25em b}\kern-0.8em\TeX}}}

\setcopyright{acmcopyright}
\copyrightyear{2024}
\acmYear{2024}
\acmDOI{XXXXXXX.XXXXXXX}
\acmConference[RobustRecSys @ RecSys '24]{RobustRecsys workshop at RecSys}{Oct. 14--18,
  2024}{Bari, Italy}
\acmBooktitle{RobustRecsys workshop at the 18th ACM Conference on Recommender Systems (RecSys '24), October 14--18, 2024, Bari, Italy}
\acmPrice{15.00}
\acmISBN{978-1-4503-XXXX-X/18/06}

\begin{document}
\title[Robust Training Objectives Improve Embedding-based Retrieval \\in Industrial Recommendation Systems]{Robust Training Objectives Improve Embedding-based Retrieval in Industrial Recommendation Systems}
\author[Kolodner et al.]{Matthew Kolodner, Mingxuan Ju, Zihao Fan, Tong Zhao, Elham Ghazizadeh, Yan Wu, Neil Shah, Yozen Liu}
\affiliation{%
  \institution{Snap, Inc.}
  \streetaddress{2772 Donald Douglas Loop N}
  \city{Santa Monica}
  \state{CA}
  \country{USA}
  \postcode{90405}
}
\email{{mkolodner, mju, zfan3, tong, eghazizadeh, ywu, nshah, yliu2}@snap.com}

\begin{abstract}
    Improving recommendation systems (RS) can greatly enhance the user experience across many domains, such as social media. Many RS utilize embedding-based retrieval (EBR) approaches to retrieve candidates for recommendation. In an EBR system, the embedding quality is key. According to recent literature, self-supervised multitask learning (SSMTL) has showed strong performance on academic benchmarks in embedding learning and resulted in an overall improvement in multiple downstream tasks, demonstrating a larger resilience to the adverse conditions between each downstream task and thereby increased robustness and task generalization ability through the training objective. However, whether or not the success of SSMTL in academia as a robust training objectives translates to large-scale (i.e., over hundreds of million users and interactions in-between) industrial RS still requires verification. Simply adopting academic setups in industrial RS might entail two issues. Firstly, many self-supervised objectives require data augmentations (e.g., embedding masking/corruption) over a large portion of users and items, which is prohibitively expensive in industrial RS. Furthermore, some self-supervised objectives might not align with the recommendation task, which might lead to redundant computational overheads or negative transfer. In light of these two challenges, we evaluate using a robust training objective, specifically SSMTL, through a large-scale friend recommendation system on a social media platform in the tech sector, identifying whether this increase in robustness can work at scale in enhancing retrieval in the production setting. Through online A/B testing with SSMTL-based EBR, we observe statistically significant increases in key metrics in the friend recommendations, with up to \underline{$\mathbf{5.45}\%$} improvements in new friends made and \underline{$\mathbf{1.91}\%$} improvements in new friends made with cold-start users. Besides, with a dedicated case study, the benefits of robust training objectives are demonstrated through SSMTL on large-scale graphs with gains in both retrieval and end-to-end friend recommendation.
\end{abstract}

\maketitle

\section{Introduction}
Recommendation systems (RS) have become a crucial component for user experience
~\cite{li2023recent, sun2024survey}. 
Most industrial RS explore a two-stage~process~\cite{10.1145/2959100.2959190}. 
During the first stage (i.e., the \textit{retrieval} phase), among hundreds of millions of candidate users/items, the RS usually utilizes several models optimized for recall to select a small set of candidate users/items (e.g., 1,000 candidates).
Whereas during the second stage (i.e., the \textit{ranking} phase), within the candidate subset, the RS can explore complicated expensive models that are optimized for precision to select top $K$ candidates for the final recommendation.
Such two-stage process enables recommendation over large quantities of possible users/items and allows for greater flexibility towards key recommendation metrics. 

In this two-stage scheme, the retrieval stage is especially important, as it acts as the bottleneck for possible candidates provided to the ranker in the second stage. 
One common approach~\cite{DBLP:journals/corr/abs-2006-11632, gan2023binary} for the retrieval step is to leverage embedding-based retrieval (EBR). 
Specifically, EBR learns embeddings for all users and items as vectors in a low-dimensional latent space. 
These embeddings are learned in a way such that the distance between them is reflective of their similarity, with more similar items being closer together in the latent space. 
As a result, candidates can be retrieved through a nearest-neighbor search across the latent space. In practice, this is done using an approximate nearest neighbor methods optimized for large-scale retrieval, such as FAISS~\cite{DBLP:journals/corr/JohnsonDJ17} and HNSW~\cite{DBLP:journals/corr/MalkovY16}.

Many methods \cite{zhang2023divide, 1167344, jha2023unified, peng2023embeddingbased} have been proposed for generating high-quality embeddings for EBR, which lead to more relevant candidates and improved metrics after the end-to-end recommendation. 
In this work, we specifically focus on the friend recommendation EBR setting, where vast amounts of topological information relating users are readily available. 
Recent works \cite{10.1145/3539618.3591848, DBLP:journals/corr/abs-1806-01973,kung2024improving} have shown that including this relational information can improve the embedding quality. The relational information is commonly modeled with graph neural networks (GNNs), producing embeddings that leverage neighbor information in graphs, such as co-friend relationships.
For graph-aware EBR in particular, link prediction has seen success for generating high-quality embeddings \cite{LI2021516}, where we look to predict the presence of an edge between a query node and set of candidate nodes. 

While link prediction is effective in learning nuanced similarities and distinctions between candidates, there are several other self-supervised graph learning philosophies that can provide high-quality embeddings, such as mutual information maximization~\cite{oord2018representation}, generative reconstruction~\cite{he2022masked}, or whitening decorrelation~\cite{ermolov2021whitening}. Based on these general philosophies, many graph-based approaches have been proposed and used to learning embeddings directly, achieving desirable properties of embeddings without requiring explicit labels. Recently, Ju et al.~\cite{ju2023multitask} evaluated combining these self-supervised learning approaches with link prediction in a multitask (MTL) setting, demonstrating a larger resilience to the adverse conditions between each downstream task and thereby increased robustness and generalization ability through the training objective

However, whether or not using SSMTL in academia as a robust training objective translates to large-scale (i.e., over hundreds of millions of users and interactions in-between) industrial RSs still requires verification. 
Simply adopting academic setups in industrial RSs might result in several issues. 
Firstly, many self-supervised objectives require data augmentations (e.g., embedding masking/corruption) over a large portion of users and items, which is prohibitively expensive in industrial RSs. 
Furthermore, some self-supervised objectives might not align with the recommendation task, which might lead to redundant computational overheads or negative transfer ~\cite{TorreyShavlik2010}, a phenomenon where performance can worsen as a result of the complexity and potentially opposing nature of the various tasks.

In this work, we investigate whether robust SSMTL training objectives are able to improve the link prediction retrieval performance on large-scale graphs with over hundreds of millions of nodes and edges. 
Specifically, we look to find what combination of SSL approaches can improve overall robustness and thereby augment retrieval through complementary yet disjoint information. 
In our experiments, we find two SSL approaches, based on philosophies from whitening decorrelation (e.g., Canonical Correlation Analysis \cite{DBLP:journals/corr/abs-2106-12484}) and generative reconstruction (e.g., Masked Autoencoders \cite{hou2022graphmae}), that are able to augment the performance of link prediction without negative transfer. 
We deploy the proposed framework on an industrial large-scale friend recommendation system to a community of hundreds of millions of users.
In the online A/B testing, we observe significant improvements in key metrics like new friends made, especially with cold-start users on the platform. Our contributions are summarized as follows:
    \begin{itemize}
        \item We demonstrate the effectiveness of robust training objectives such as SSMTL in a large-scale industrial recommendation system.
        \item We conduct an online study of SSMTL on a massive real-world recommendation system, and observe a statistically significant increase in key metrics, with up to \underline{$\mathbf{5.45}\%$} improvements in new friends made and \underline{$\mathbf{1.91}\%$} improvements in new friends made with cold-start users.
    \end{itemize}
\definecolor{CCA}{HTML}{b6d7a8}
\definecolor{MAE}{HTML}{ea9999}

\begin{figure*}
   \centering
   \includegraphics[scale=0.9, width=\textwidth]{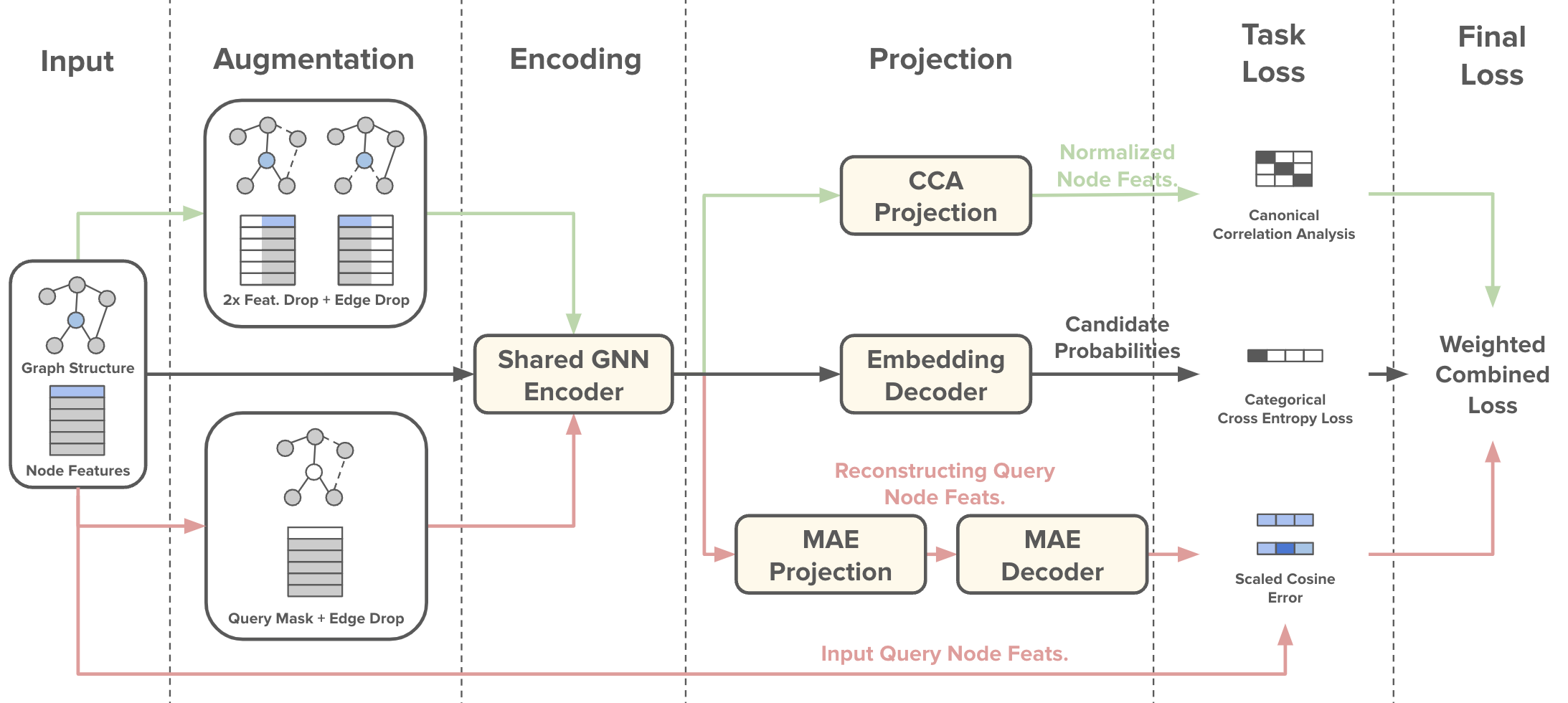} % Specify the image file path and desired width
   
% In our proposed SSMTL framework, we combine the CCA and MAE SSL methods in our embedding generation scheme for EBR.
   \caption{In our proposed SSMTL framework, we combine the \textcolor{CCA}{CCA} and \textcolor{MAE}{MAE} SSL methods with the retrieval task in our embedding generation scheme for EBR. \textcolor{CCA}{CCA} looks to maximize the correlation of two augmented views of the input subgraph while decorrelating features of a single view. \textcolor{MAE}{MAE} seeks to reconstruct the query user nodes after being propagated through the GNN encoder backbone. Finally, the retrieval task seeks to predict which candidates share a link with the query user using a categorical cross entropy loss. The loss of each subtask is weighted and summed to measure the final loss. Embeddings can be generated through the GNN encoder for EBR.}
   \label{fig:your_label}
\end{figure*}

\section{Background}
\subsection{Graph-Aware Embedding-based Retrieval}

%- Talks about definition of EBR in general
%- How graph aware EBR work (can check Pau's paper or the previous EBR paper
%- a short mention that we apply this to the friend rec sys at Snap 

In a two-stage recommendation system with a retrieval then ranking phase, the retrieval phase plays an important role filtering out the most relevant candidates to lighten the load of the ranker. Since the ranking result is largely dependent on items retrieved in the retrieval phase, a good quality retrieval model can drastically improve the final ranking. Embedding based retrieval (EBR) is a method that's recently adopted and deployed in many content, product, and friend recommendation systems\cite{DBLP:journals/corr/abs-2006-11632, 45530, TK:2021,10.1145/3539618.3591848}, and proved to achieve superior results. EBR transform users and items into embeddings, changing the retrieval problem into a nearest-neighbor search problem in a low-dimensional latent space. These embeddings can be determined in advance and indexed using an approximate nearest neighbor search such as FAISS \cite{DBLP:journals/corr/JohnsonDJ17} and HNSW \cite{DBLP:journals/corr/MalkovY16} in order to retrieve the top-$k$ most relevant items efficiently at serving.

When applying EBR to RS problems, the quality of embeddings is of upmost importance. In this paper, we use a friend recommendation system as our subject. In scenarios like friend recommendation where vast amounts of topological information relating users and items is readily available, these embeddings can be augmented with GNNs. Previous work showed that EBR for friend recommendation systems see benefits leveraging graph-aware embeddings\cite{10.1145/3539618.3591848}. In this setting, nodes would contain individual user features while edges map to user-user interactions. This approach compliments commonly used graph traversal approaches (eg. friend-of-friend (FoF) \cite{Newman_2001}), allowing for retrieval of candidates from any number of hops away from the target.

Here we describe GNNs for generating graph-aware embeddings for EBR. 
GNNs have demonstrated state-of-the-art performance in many problems containing rich topological information within the graph data \cite{DBLP:journals/corr/abs-1812-08434}, such as recommendation and forecasting. Formally, we define $G = (\mathcal{V}, \mathcal{E}, X)$, where $\mathcal{V}$ is the set of $n$ nodes ($|\mathcal{V}| = n)$, $\mathcal{E}$ is the set of edges ($\mathcal{E} \in \mathcal{V} \subseteq \mathcal{V}$), and $X$ is a feature matrix of dimension $d$ where $X \in \mathbb{R}^{n \times d}$. Many modern GNNs also employ a message-passing structure, consisting of an aggregation (AGG) and update (UPD) function. The goal of this paradigm is for nodes to receive information from their neighbors, collecting messages using its AGG function before updating their own messages with the UPD function, both of which are learnable and permutation-invariant. For some node $u$ at layer $k$, the next message-passing layer can be written as 
% Remember to bold equations for matrix/vector
\begin{equation}
    \mathbf{h}_u^{(k+1)} = \text{UPD}^{(k)}\left( \mathbf{h}_u^{(k)}, \text{AGG}^{(k)}\left( \{\mathbf{h}_v^{(k)}, \forall v \in \mathcal{N}(u)\} \right) \right)
\end{equation}

where $\mathcal{N}(u)$ is the neighborhood nodes of node $u$. Different message-passing GNN models use different combinations of AGG and UPD functions. An example of a more complex GNN, Graph-attention networks (GATs) \cite{veličković2018graph}, use an attention mechanism for each pair of nodes $i$ and $j$ 

\begin{equation}
    \alpha_{ij} = \text{softmax}_j \left( f_{\text{att}} \left( \mathbf{W}h_i, \mathbf{W}h_j \right) \right)
\end{equation}
where $\mathbf{W}$ is a linear transformation applied to every node and $f_{\text{att}}$ is the attention function parameterized by a weight vector and a non-linearity function. The AGG function is then a attention-weighted sum of its neighbors features while the UPD function is implicitly defined in $\mathbf{W}$ and the non-linearity function. 
Typically, to generate graph-aware embeddings from GNNs, a margin based ranking loss\cite{DBLP:journals/corr/abs-1806-01973, 10.1145/3539618.3591848} or contrastive\cite{DBLP:journals/corr/abs-2101-01317} loss can be used, to encourage items that are closer in the graph to be closer in the embedding space. 

\subsection{Multitask Learning}
Multitask learning (MTL) is an approach in machine learning where a model is trained simultaneously on several tasks. MTL has been extensively explored in recommendation as a way to improve key metrics \cite{inproceedings1, 10.1145/3219819.3220007, ma2018entire, 10.1145/3383313.3412236}. Thus, the core idea behind multitask learning is to improve the robustness of the model by leveraging the domain-specific information contained in the training signals of related tasks \cite{NIPS2006_0afa92fc, Caruana1997MultitaskLearning}. Hard parameter sharing, one of the most fundamental forms of MTL, uses a shared representation which then branches into multiple heads capable of learning task-specific information \cite{guo2020learning, DBLP:journals/corr/abs-1911-12423, vandenhende2020branched}.

For graph-aware EBR in particular, self-supervised multitask learning (SSMTL) has been proposed as a new approach to MTL, optimizing the embeddings directly to achieve desirable embedding properties without the use of positive or negative labels. In this setting, we combine several self-supervised learning (SSL) methods with a downstream retrieval task to learn both direct and indirect embedding features. Recent work \cite{ju2023multitask} has shown that SSMTL can lead to improved task generalization and embedding quality on several academic benchmarks through the increasingly robust training objective.  However, many of the SSL approaches used are constrained to the assumption that global graph information can be inferred within the graph structure. This is not valid in the large-scale recommendation setting, where graphs are constrained to some $K$-hop around a query user in order to fit in memory. As a result, many of these SSL methods may lead to negative transfer due to SSL task conflict with the target link prediction task, and there remains work to be done to investigate which methods perform best in this large-scale setting.
\section{Self-Supervised Multitask Learning for EBR}

% To evaluate the task compatibility of SSMTL approaches on large-scale EBR recommendation systems, we conduct a thorough investigation of different combinations of SSL methods built over GNN architectures. 
In the following sections, we describe details of the SSL methods used in our SSMTL approach, our experiment set up and results, highlighting the benefits and impact of including SSMTL based embedding in EBR for large-scale industrial recommendation systems.

% From our experiments, we report the setting that leads to the largest improvement on the link prediction task, resulting in better embedding quality for EBR. We then highlight our solution as a general framework for transferring SSMTL benefits from academic benchmarks to the large-scale recommendation setting. 

\subsection{Self-Supervised Learning Methods}

We identify two self-supervised learning approaches that are scalable and lead to improvements in the large-scale recommendation setting through a more robust training objective. \\

\textbf{Canonical Correlation Analysis.}
Based on work from \cite{DBLP:journals/corr/abs-2106-12484}, Canonical Correlation Analysis (CCA) deploys a non-contrastive, non-discriminitive SSL method to train the GNN. The self-supervised training objective is described in Equation \ref{eq:1}. First, given a subgraph with $n$ nodes, two augmented views of the subgraph are created and fed through the GNN, producing $\mathbf{Z_A}$ and $\mathbf{Z_B}$ where $\mathbf{Z_A}, \mathbf{Z_B} \in \mathbb{R}^{n \times k}$. Each of these embeddings are fed through a task-specific head, and then are normalized so that each feature has $0$ mean and $\frac{1}{\sqrt{n}}$ standard deviation, resulting in $\mathbf{\tilde{Z}_A}$ and $\mathbf{\tilde{Z}_B}$. The loss is then computed from Equation \ref{eq:1}. The first term in the equation seeks to minimize the distance of the same nodes between the two views. The second term enforces that the feature-wise covariance of all nodes is equal to the identity matrix. 

\begin{equation}
    \mathcal{L_{\text{CCA}}} = \left\| \tilde{\mathbf{Z}}_A - \tilde{\mathbf{Z}}_B \right\|_F^2 + \lambda \left( \left\| \tilde{\mathbf{Z}}_A^T \tilde{\mathbf{Z}}_A - \mathbf{I} \right\|_F^2 + \left\| \tilde{\mathbf{Z}}_B^T \tilde{\mathbf{Z}}_B - \mathbf{I} \right\|_F^2 \right)
    \label{eq:1}
\end{equation}
\textbf{Masked Autoencoders.}
Based on work from \cite{hou2022graphmae}, this approach leverages a graph masked autoencoder (MAE) that focuses on feature reconstruction. First, an augmented view of the subgraph is created and the features of the query users are masked out. This augmented graph is then fed through the GNN and a task-specific head. The features of the query users are then re-masked and passed through a graph convolution layer. As described in Equation \ref{eq:2}, for all masked nodes $\mathcal{V}$, the final loss is equal to the average of the scaled cosine error between the original features $\mathbf{X}$ and generated features $\mathbf{Z}$. This approach only relies on the local neighborhood surrounding the query node, making it a good option for large-scale SSMTL. 
\begin{equation}
    \mathcal{L}_{\text{MAE}} = \frac{1}{|\mathcal{V}|} \sum_{v_i \in \mathcal{V}} \left(1 - \frac{\mathbf{x}_i^T \mathbf{z}_i}{\|\mathbf{x}_i\| \cdot \|\mathbf{z}_i\|}\right)^y, \quad y \geq 1
    \label{eq:2}
\end{equation}
We note that these two approaches both utilize non-contrastive methods. While experimenting with different SSL tasks, we find that contrastive SSL approaches do not perform very well in the production setting due to their assumption that global information is readily available in the original and augmented graphs. This is not necessarily true for large-scale recommendation, where subgraphs are constrained to the K-hop neighborhood surrounding each query node. 

\subsection{Experimental Setup}
\subsubsection{Problem Breakdown}
We evaluate the SSMTL as a robust training objective on an industrial friend recommendation system with hundreds of millions of users and connections. To handle this scale of training, we sample subgraphs containing the $k$-hop neighborhood around each query user. Following training, the embeddings for EBR can be via propagation through the encoder backbone. 

\subsubsection{Retrieval Baseline}
The baseline model uses a supervised single-task setup for embedding-based retrieval. We use a GAT as the GNN encoder backbone to obtain embeddings for the query user and each candidate, producing a candidate embedding matrix $\mathbf{z}$. We can then compute the dot product between the query user and each candidate and apply Softmax to generate the logits. We then calculate the Categorical Cross Entropy Loss with the true labels $\mathbf{y}$ across the $N=2$ classes and $M$ candidates, outlined in Equation \ref{eq:3}. 
\begin{equation}
    \mathcal{L}_{\text{retrieval}} = -\sum_{i=1}^{N} \sum_{j=1}^{M} y_{ij} \log\left(\frac{e^{z_{ij}}}{\sum_{k=1}^{M} e^{z_{ik}}}\right)
    \label{eq:3}
\end{equation}
\subsubsection{SSMTL Implementation Details}
In our SSMTL approach, we use both CCA and MAE in combination with the retrieval baseline as the training objectives. All three methods share the same GAT GNN backbone. The augmented views for CCA and MAE occur separately, with CCA performing edge and feature drop augmentations while MAE performs edge drop and query node masking. The task-specific head for CCA is a Linear-ReLU-Linear block while the task-specific head for MAE is one linear layer. The final loss with SSMTL is a weighted sum of the losses.

\begin{equation}
\mathcal{L_{\text{combined}}} = \alpha \mathcal{L}_{\text{retrieval}} + \beta \mathcal{L}_{\text{CCA}} + \gamma \mathcal{L}_{\text{MAE}}
\end{equation}
where $\alpha$ is the weight for the retrieval loss, $\beta$ is the weight of the CCA loss, and $\gamma$ is the weight of the MAE loss. In practice, we observed best performance when the retrieval weight was several orders of magnitude larger than the other loss weights.

%\subsubsection{Offline Evaluation}
%To evaluate the effectiveness of SSMTL in an isolated retrieval setting, we conduct offline experiments where we look to predict links for target users. We evaluate the performance of our models using mean-reciprocal rank (MRR), as shown in Equation \ref{eq:MRR}, where $|Q|$ is the number of queries and $\text{rank}_i$ is the rank of the correct candidate. 

%\begin{equation}
%    \label{eq:MRR}
%    MRR = \frac{1}{|Q|} \sum_{i=1}^{|Q|} \frac{1}{\text{rank}_i}
%\end{equation}

% Include Clark's results
\subsection{Results}
%In the offline setting, we observed a $1\%$ improvement in MRR after adding the canonical correlation analysis and masked graph autoencoder self-supervised tasks compared to the baseline. 
We evaluated the effectiveness of SSMTL for end-to-end friend recommendation with online A/B testing. The control group used candidates retrieved from the production model trained with retrieval baseline, while the treatment group instead used candidates retrieved with the new robust training objective in the SSMTL setting, specifically combining the previous retrieval loss with whitening decorrelation and generative reconstruction objectives.

In the A/B experimental results, we saw statistically significant improvements across several friend recommendation metrics. Specifically, we observed up to \textbf{\textit{5.45\%}} improvements in new friends made and \textbf{\textit{+1.91\%}} new friends made with low-degree users in various markets. Overall, from these results, we see that SSMTL is able to provide improved recommendation compared with the single-task setting, in particular helping with candidate generation for low-degree users.

\section{Conclusion}
In this paper, we evaluate the effectiveness of a robust self-supervised multitask learning objective in embedding-based retrieval. Through online evaluation, we demonstrate that self-supervised methods used in a multi task setting are able to augment the performance of the underlying retrieval task on the scale of over 800 million nodes and edges, providing complementary yet disjoint information to enhance the embedding quality. We observe statistically significant gains in the number of friendships made for both high and low degree users. 

% For future work, we are looking forward to continuing evaluation of different tasks and loss combinations. We will also investigate incorporating more tasks with different types of data input. 

\bibliographystyle{unsrt}
\bibliography{references}

\end{document}